\documentclass[twocolumn,showpacs,preprintnumbers,amsmath,amssymb]{revtex4}
\usepackage{epsf}
\begin{document}
\title{Exploiting quantum parallelism of entanglement for a complete
experimental quantum characterization of a single qubit device}
\author{Francesco De Martini}
\email{francesco.demartini@uniroma1.it}
\author{Andrea Mazzei}
\email{andrea.mazzei@uniroma1.it}
\author{Marco Ricci} 
\email{marco.ricci@uniroma1.it}
\affiliation{Istituto Nazionale di Fisica della Materia, Dipartimento
di Fisica, Universit\`a ''La Sapienza'', Roma, 00185 Italy}
\homepage{http://atesit.phys.uniroma1.it}
\author{Giacomo Mauro D'Ariano}
\email{dariano@unipv.it}
\affiliation{{\em Quantum Optics and Information Group},
Istituto Nazionale di Fisica della Materia,
Unit\`a di Pavia}
\homepage{http://www.qubit.it}
\affiliation{Dipartimento di Fisica ``A. Volta'', via Bassi 6, I-27100 Pavia, Italy}
\altaffiliation[Also at ]{Department of Electrical and Computer
Engineering, Northwestern University, Evanston, IL  60208}
\date{\today}

\begin{abstract}
We present the first full experimental quantum tomographic
characterization of a single-qubit device achieved with a single
entangled input state. The entangled input state plays the role of
all possible input states in quantum parallel on the tested device.
The method can be trivially extended to any $n$-qubits device by just
replicating the whole experimental setup $n$ times.
\end{abstract}
\maketitle
\date{October 30, 2002}
\maketitle
The new field of Quantum Information\cite{nielsen_book} has recently
opened the way to realize radically new processing devices, with the
possibility of  tremendous speedups of complex computational
tasks, and of cryptographic communications guaranteed by the laws of
physics. Among the many problems posed by the new information
technology there is the need of making a complete experimental
characterization of the 
functioning of the new quantum devices. As shown recently in
Ref. \cite{darianolopresti}, quantum mechanics provides us with the
perfect tool to achieve the task easily and efficiently: this is the
so called {\em quantum entanglement}, the basis of the quantum
parallelism of future computers. In this letter we present the first
full experimental quantum characterization of a single-qubit device,
based on this method. Since the method can be easily extended to any
$n$-qubit device, the present experiment represents  a first test of the
feasibility and of the experimental limits of the new general
tomographic method. 

How we characterize the operation of a device? In quantum
mechanics the evolution of the state is completely described by the so
called {\em quantum operation}\cite{kraus} of the device, that here we
will denote by ${\mathrm E}$. More precisely, the output state
$\rho_{out}$ is given by the quantum operation ${\mathrm E}$ applied to
the input state $\rho_{in}$ as follows 
\begin{equation}
\rho_{out}=\frac{{\mathrm E}\left( \rho_{in}\right)
}{\mbox{Tr}\left[{\mathrm E}\left( \rho_{in}\right) \right]}.\label{Sred}
\end{equation}
The normalization constant $\mbox{Tr}\left[{\mathrm E}\left(
\rho_{in}\right)\right]$ in Eq. (\ref{Sred}) is also the probability of occurrence of the
transformation ${\mathrm E}$ (e. g. when there are other possible alternatives,
such as when we consider the state transformation due to a measuring
device for a given outcome).  Therefore, apart from a normalization
factor, the quantum evolution is always linear, with the quantum
operation playing the role of the so-called {\em transfer matrix} of
the device, a mathematical tool very popular in optics and electrical engineering.   
\par Now the problem is: how to reconstruct the form of ${\mathrm E}$ experimentally?
Since ${\mathrm E}$ is essentially a transfer matrix for a linear system, one
would be tempted to adopt the conventional method of running an {\em
orthogonal basis} $|n\rangle$ of inputs and measuring the corresponding
outputs by {\em quantum tomography}\cite{varenna}. However, since
states are actually operators---not vectors---as a consequence of the
polarization identity in order to get all off-diagonal (complex) matrix
elements of the state one actually needs to run not the basis 
itself, but all the linear combinations of its vectors
$2^{-\frac12}\left(|n^\prime\rangle +\kappa|n^{\prime\prime
}\rangle\right) $, with $\kappa=\pm 1,\pm i$. In the following
we will call such sets of states {\em faithful}, since they are
sufficient to make a complete tomography of a quantum operation.
This method is essentially the {\em quantum process tomography} given in 
Ref. \cite{nielsen_book}, which was experimentally demonstrated in
NMR \cite{NMR} and recently in quantum optics \cite{kwiat}. The main
problem with such method is the fact that 
in most practical situations faithful sets of input states are not
feasible in the lab. For example, for continuous-variables 
process tomography in the Fock representation, the states 
$| n^\prime\rangle$ and $| n^{\prime\prime}\rangle$ would be photon
number states, and achieving their superpositions will remain a dream
for experimentalists for many years. But here the {\em quantum
parallelism} of entanglement comes to help us, with a single input entangled state
that is equivalent to running all possible input states in parallel
\cite{darianolopresti}. Thus, we don't need to prepare a complete set
of states, but just a single entangled one, a state commonly available
in any modern quantum optical laboratory.
\par
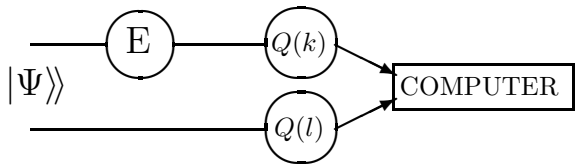
\begin{figure}[hbt]
\begin{center}
\setlength{\unitlength}{700sp}%
\begingroup\makeatletter\ifx\SetFigFont\undefined%
\gdef\SetFigFont#1#2#3#4#5{%
  \reset@font\fontsize{#1}{#2pt}%
  \fontfamily{#3}\fontseries{#4}\fontshape{#5}%
  \selectfont}%
\fi\endgroup%
\begin{picture}(19506,5446)(1549,-5109)
\thicklines
\put(1801,-3961){\line( 1, 0){8400}}
\put(12601,-961){\vector( 2,-1){2160}}
\put(12601,-3961){\vector( 2, 1){2160}}
\put(1801,-961){\line( 1, 0){2700}}
\put(6901,-961){\line( 1, 0){3300}}
\put(14701,-3211){\framebox(6300,1500){}}
\put(17701,-2761){\makebox(0,0)[b]{\smash{\SetFigFont{10}{49.2}{\rmdefault}{\mddefault}{\updefault}COMPUTER}}}
\put(11326,-1186){\makebox(0,0)[b]{\smash{\SetFigFont{10}{49.2}{\rmdefault}{\mddefault}{\updefault}$Q(k)$}}}
\put(11326,-4186){\makebox(0,0)[b]{\smash{\SetFigFont{10}{49.2}{\rmdefault}{\mddefault}{\updefault}$Q(l)$}}}
\put(5626,-1261){\makebox(0,0)[b]{\smash{\SetFigFont{15}{49.2}{\rmdefault}{\mddefault}{\updefault}${\mathrm E}$}}}
\put(2026,-2761){\makebox(0,0)[b]{\smash{\SetFigFont{15}{49.2}{\rmdefault}{\mddefault}{\updefault}$|\Psi\rangle\!\rangle$}}}
\put(5701,-961){\oval(2372,2372)}
\put(11401,-886){\oval(2500,2500)}
\put(11401,-3886){\oval(2500,2500)}
\end{picture}
\end{center}
\caption{General experimental scheme of the method for the tomographic
estimation of the quantum operation ${\mathrm E}$ of a single qubit device. Two identical
quantum systems---two optical beams in the present experiment---are
prepared in an entangled state $|\Psi\rangle\!\rangle$. One of the two
systems undergoes the quantum operation ${\mathrm E}$, whereas the other
is left untouched. At the output one makes a quantum 
tomographic estimation, by measuring jointly two observables (each for
each beam) from a {\em quorum} $\{Q(l)\}$. In the present
experiment the quorum is represented by the 
set of Pauli operators $\sigma_x$, $\sigma_y$ and $\sigma_z$.}\label{gener}\end{figure}

 In the following we will use the double-ket notation
$|\Psi\rangle\!\rangle\in{\cal H}\otimes{\cal H}$ 
to denote bipartite states corresponding to the matrix $\Psi_{nm}$ of
coefficients on fixed given orthonormal basis
$|n\rangle\otimes|m\rangle\equiv|nm\rangle$ of ${\cal H}\otimes{\cal H}$ 
\begin{equation}
|\Psi\rangle\!\rangle =\sum_{nm}\Psi_{nm}| nm\rangle.\label{PSI}
\end{equation}
In our optical implementation the entangled systems consist of two
single-mode optical beams, and the Hilbert space is
two-dimensional, since we will consider only single-photon polarization
states.  The key feature of the method implies that only one of the
two systems is input into the {\em unknown} transformation ${\mathrm E}$,
whereas the other is left untouched, as in Fig. \ref{gener}. This setup leads to the output
state $R_{out}$, which in tensor notation writes as follows 
\begin{equation}
R_{out}={\mathrm E}\otimes{\mathrm I}(|\Psi\rangle\!\rangle
\langle\!\langle \Psi|),
\end{equation}
where ${\mathrm I}$ denotes the identical operation. It is a result of linear algebra
that $R_{out}$ is in one-to-one correspondence with the quantum operation ${\mathrm E}$,
as long as the state $|\Psi\rangle\!\rangle$ is {\em full-rank}, i. e. its
matrix $\Psi$ is invertible. This is the case of a so-called {\em
maximally entangled} state, where the matrix $\Psi$ is proportional to
a unitary one. Full-rank entangled states can be easily generated by
spontaneous parametric down conversion of vacuum, as in the
experiment reported here. Note that even when a faithful set of input
states is available in the lab---which is actually true in our case of
single-photon polarization states---nevertheless a single faithful
entangled state can be much more efficient and more practical.
As a matter of fact, in practice, generation of  single-photon polarization states
relays anyway on entanglement, and, as we will see in the following,
the present method uses all experimental data much more efficiently
than the conventional quantum process tomography
\cite{nielsen_book,kwiat}.

\begin{figure}[hbt]\begin{center}
\epsfxsize=.8\hsize\leavevmode\epsffile{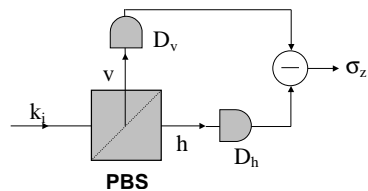}
\end{center}
\caption{Pauli-matrix $\sigma_z$ measurement apparatus for
photon-polarization qubits inserted at the end of each optical
beam. The beam is split by a polarizing beam splitters (PBS) into its
horizontal and vertical components, which are separately detected and
recorded with plus and minus sign, respectively. For measuring the
other two Pauli matrices $\sigma_x$ and $\sigma_y$, the PBS is
preceded by a suitably oriented $\frac{\lambda}{2}$ and
$\frac{\lambda}{4}$ wave-plate, 
respectively (see text).}\label{paulimeas}\end{figure}

Now, how to characterize the entangled state R$_{out}$ at the output?
For this purpose a technique for the full determination of the quantum
state has been introduced and developed since 1994. The method named
{\em Quantum Tomography} \cite{varenna} has been initially introduced
for the state of a single-mode of radiation, the so called {\em
Homodyne Tomography}, and thereafter it has been generalized to any
quantum system. The basis of the method is the measurement of a
suitably complete set of observables called {\em quorum}. In our case
we need to measure jointly two quora of observables on the entangled
qubit, here the quorum being the Pauli matrices $\sigma_x$, 
$\sigma_y$, $\sigma_z$. The qubit is encoded on polarization of single
photons, as follows  
\begin{equation}
|0\rangle\doteq h^\dag|\Omega\rangle,\qquad
|1\rangle\doteq v^\dag|\Omega\rangle,\label{logic}
\end{equation}
where $|\Omega\rangle$ denotes the e. m. vacuum, and $h$, $h^\dag$ and
$v$, $v^\dag$ the annihilation and creation operators of the
horizontally and vertically polarized modes associated to a fixed
wave-vector ${\mathbf k}$, respectively. In synthesis,
Eq. (\ref{logic}) means that the ''logical zero'' is encoded on a
single horizontally polarized photon,  whereas the ''logical one'' is
encoded on a vertically polarized photon. In the present 
representation, the Pauli matrices write as follows 
\begin{equation}
\sigma_x=h^{\dagger }v+v^{\dagger }h,\;
\sigma_y=i\left( h^{\dagger }v-v^{\dagger }h\right),\;
\sigma_z=h^{\dagger }h-v^{\dagger }v.\!\label{pauli}
\end{equation}
According to Eq. (\ref{pauli}), the $\sigma_z$-photo-detector is
simply achieved as in Fig. \ref{paulimeas}. 
In order to understand how to design detectors for $\sigma_y$ and
$\sigma_z$ we still need some simple algebra for wave-plates. 
The ring of Pauli matrices is completed by including the ``identity''
$\sigma_0=h^\dagger h+v^\dagger v$, corresponding to single-photons
states. In the following we use the popular relativistic conventions, 
denoting by $\vec{\sigma}$ the three-vector of operators $\vec{\sigma}$ =
$\left(\sigma_1,\sigma_2,\sigma_3\right) $ and by $\sigma $ the 
tetra-vector $\sigma $ = $\left( \sigma_{0},\sigma_1,\sigma_2,\sigma
_3\right) $, and use Greek indices for three-vectors components $\alpha
=1, 2, 3$ (or $\alpha=x, y, z$), and Latin indices for tetra-vector
components $i = 0, 1, 2, 3$.
\par A wave-plate changes the two radiation modes according to the matrix
transformation
\begin{equation}
\begin{bmatrix}h \\ v\end{bmatrix}
\longrightarrow
{\mathbb W}_{\phi,\theta}
\begin{bmatrix}h \\ v\end{bmatrix}
\label{Wtrans}
\end{equation}
where the matrix ${\mathbb W}_{\phi,\theta}$ is given by
\begin{equation}
{\mathbb W}_{\phi ,\theta}=
\begin{bmatrix}
z_{+}+cz_{-} & sz_{-} \\ sz_{-} & z_{+}-cz_{-}
\end{bmatrix}
\end{equation}
where $s=\sin 2\theta$, $c=\cos 2\theta$, $z_{\pm }=\frac12\left( 1\pm
e^{i\phi }\right)$, $\theta $ is the wave-plate orientation angle
around the wave-vector ${\mathbf k}$,  $\phi =2\pi\delta/\lambda$, 
$\lambda $ is the wave-length, and $\delta$ is the length of the optical path through
the plate. Special cases are the $\frac{\lambda}{4}$ wave-plate which can be used
with $\theta =\pi/4$ to give the right $c_+$ and left $c_-$
circularly polarized modes $c_{\pm}=e^{i\pi /4} 2^{-\frac{1}{2}} (\pm
h+iv)$, and the $\frac{\lambda}{2}$ wave-plate which can be used to give the diagonal
linearly polarized modes $d_{\pm}=2^{-\frac{1}{2}}(h\pm v)$. On the
Pauli operator vector $\vec\sigma$ the mode transformation due to a
wave-plate writes
\begin{equation}
\vec{\sigma} \longrightarrow {\mathbb R}_{\phi,\theta}\vec{\sigma},\label{10}
\end{equation}
with rotation matrix
\begin{equation}
{\mathbb R}_{\phi,\theta}=\begin{bmatrix}
s^2+c^2\cos \phi & -c\cos \phi & sc(1-\cos \phi ) \\
c\sin \phi & \cos \phi & -s\sin \phi \\
sc(1-\cos \phi ) & s\sin \phi & c^2+s^2\cos \phi
\end{bmatrix}.\label{matrixtrans}
\end{equation}
From Eq. (\ref{matrixtrans}) we can see that a $\sigma_x$-detector can
be obtained by preceding the $\sigma_z$-detector with a
$\frac{\lambda}{2}$ wave-p1ate oriented at $\theta=\pi/8$, whereas a
$\sigma_y$-detector is obtained by preceding the $\sigma_z$-detector
with a $\frac{\lambda}{4}$ wave-plate oriented at $\theta=\pi/4$. When
collecting data at a $\sigma_\alpha$-detector, we will denote by
$s_\alpha=\pm 1$ the corresponding random outcome, $s_\alpha=+1$
corresponding to the $h$-detector flashing, and $s_\alpha=-1$
to the $v$-detector flashing.
The general experimental setup is then  given by two Pauli
detectors---for measuring $\sigma_\alpha$ and $\sigma_\beta$ for
varying $\alpha$ and $\beta$---at the output of the entangled beams, as
in Fig. \ref{gener}. The experimental data are collected in
coincidence, with two of the four photo-detectors firing, one for each
Pauli detector, thus guaranteeing that the result will be essentially
unaffected by quantum efficiency. The experimental correlations 
$\overline{s_i^{(1)}s_j^{(2)}}$ of the random outcomes $s_i^{(n)}$ 
of the detector at the $n$-th beam ($n=1,2$) on the entangled
state $|\Psi \rangle\!\rangle$ must coincide with the following
theoretical expectations
\begin{equation}\overline{s_i^{(1)}s_j^{(2)}}=
\langle\!\langle\Psi | (\sigma_i^{(1)}\otimes
\sigma_j^{(2)})|\Psi\rangle\!\rangle=\mbox{Tr}\left[
\Psi ^{+}\sigma_i\Psi \sigma_j^{\ast }\right],  \label{14}
\end{equation}
and, obviously, $\overline{s_i^{(1)}}\equiv
\overline{s_i^{(1)}s_{0}^{(2)}}$ and $\overline{s_i^{(2)}}\equiv 
\overline{s_{0}^{(1)}s_i^{(2)}}$. For maximally entangled states we
have the isotropy condition 
$\overline{s_\alpha^{(1)}}=$ $\overline{s_\alpha^{(2)}}=0$ for $\alpha
=x,y,z$. The four Bell states will correspond to the four Pauli
matrices $\sigma_j$ via a state coefficients matrix $\Psi$ given by 
$\Psi=\frac1{\sqrt2}\sigma_j$. On the other hand, when a quantum
device performing the unitary transformation $U$ is inserted in one of
the two entangled beams as in Fig. \ref{gener}, the entangled state
$|\Psi\rangle\!\rangle$ is changed to $U\otimes
I|\Psi\rangle\!\rangle$, which corresponds to the new coefficients
matrix $\Psi\to U\Psi$. In our lab we used the ``triplet''
state corresponding to $\Psi=\sigma_x/\sqrt2$, which is generated via
spontaneous parametric downconversion by an optical parametric
amplifier physically consisting of a nonlinear BBO (\ss-barium-borate)
crystal plate, 2 mm thick, cut for Type II 
phase matching and excited by a pulsed mode-locked ultraviolet laser UV
having pulse duration $\tau =140$ fsec and wavelength $\lambda_{p}$%
=397.5 nm.  The wavelength of the emitted photons is
$\lambda=795$nm. The measurement apparatus consisted of two equal 
polarizing beam splitters with output modes coupled to four equal
Si-avalanche photo-detectors SPCM-AQR14 with quantum efficiencies QE
$\simeq $ 0.42. The beams exciting the detectors are filtered by equal interference
filters within a bandwidth $\Delta\lambda=6$nm. The detector outputs
are finally analyzed by a computer. 
\par With the above apparatus  we want now to experimentally determine
the matrix elements of the state $|\Psi\rangle\!\rangle$ in
Eq. (\ref{PSI}). From the trivial identity
\begin{equation}
\langle nm|\Psi\rangle\!\rangle=\Psi_{nm},
\end{equation}
we obtain the matrix $\Psi_{nm}$ for the {\em input states} in terms of the
following ensemble averages
\begin{equation}
\Psi_{nm}=e^{i\varphi }\frac{\langle\!\langle\Psi| 01\rangle\langle
nm|\Psi\rangle\!\rangle}{\sqrt{\langle\!\langle\Psi| 01\rangle\langle 01|\Psi
\rangle\!\rangle}} \label{20}
\end{equation}
where $\exp \left( i\varphi \right) =\Psi_{01}/\left|
\Psi_{01}\right|$ is an unmeasurable phase factor. The choice of the vector $%
\left| 01\right\rangle $ is arbitrary, and it is needed only for the sake of
normalization, e. g. we could have used $\left| 10\right\rangle $ or $\left|
11\right\rangle $, instead. Using the {\em tomographic expansion} over the
four Pauli matrices \cite{varenna} we see that, via Eq. (\ref{20}),
the matrix element of the input state is obtained from 
the following experimental averages
\begin{equation}
\Psi_{nm}=\frac1{4\sqrt{p}}\sum_{ij}Q^{ij}_{nm}\overline{%
s_i^{(1)}s_j^{(2)}}
\end{equation}
where
\begin{equation}
p=\langle\!\langle\Psi|01\rangle\langle 01|\Psi\rangle\!\rangle
=\frac{1}{4}\overline{(1+s_3^{(1)})(1-s_3^{(2)})}\label{23}
\end{equation}
is the fraction of events with one $\sigma_z$-detector firing on $h$
and the other on $v$, and the matrix $Q^{ij}_{nm}$ is given by 
\begin{equation}
Q^{ij}_{nm}=\langle n| \sigma_i| 0\rangle
\langle m| \sigma_j| 1\rangle.\label{24}
\end{equation}
\begin{figure}[hbt]\begin{center}
\epsfxsize=.5\hsize
\leavevmode\epsffile{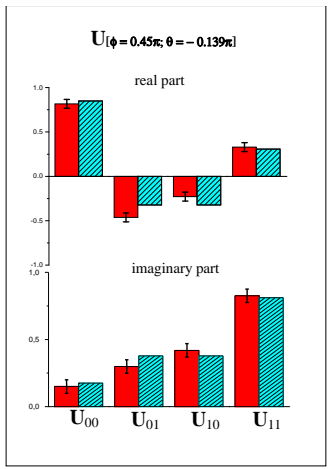}
\end{center}\vskip -.4truecm
\caption{Experimental characterization of a single optical wave-plate
with retardation phase $\phi =0.45\pi$ and orientation angle of
the optical axis respect to the laboratory horizontal direction
$\theta =-0.138\pi$.  The experimental matrix elements $U_{nm}$ of the
wave-plate are superimposed to statistical errors for 8000 events,
and compared with the theoretical  values.}\label{1wav}\end{figure}    
The unitary matrix $U_{nm}$ of the device is now obtained with the same
averaging as above, but now for the state at the output of the device
$|U\Psi\rangle\!\rangle =\left(U\otimes
I\right)|\Psi\rangle\!\rangle$, namely
\begin{equation}
\left( U\Psi \right)_{nm}=e^{i\varphi }\frac{\langle\!\langle
U\Psi| 01\rangle\langle nm|\Psi U\rangle\!\rangle}{\sqrt{\langle\!\langle
U\Psi| 01\rangle\langle 01|\Psi U\rangle\!\rangle}} \label{25}
\end{equation}
where we use again Eqs. (\ref{23}) and (\ref{24}), but now the average
expressed by Eq. (\ref{20}) is carried out over the output state $|U\Psi\rangle\!\rangle$. 
The (complex) matrix elements $U_{nm}$ are obtained from Eq. (\ref{25}) by
matrix inversion. This is of course possible since the matrix $\Psi $
is invertible, due to the maximally entangled character of $|\Psi\rangle\!\rangle$.
\par
\begin{figure}[hbt]\begin{center}\vskip .3truecm
\epsfxsize=.5\hsize\leavevmode\epsffile{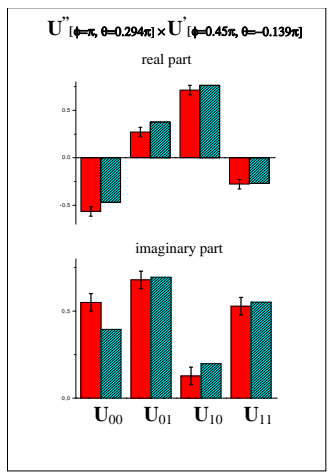}
\end{center}\vskip -.4truecm
\caption{The same experimental characterization as in Fig.
\ref{1wav}, here for a device made of a series of two optical
wave-plates: a wave-plate with $\phi=0.45\pi$ and $\theta=-0.138\pi$
followed by another wave-plate with $\phi=\pi$ and $\theta=+0.29\pi$.}\label{2wav}\end{figure}

 An experimental demonstration of the tomographic method is given in
Figures \ref{1wav} and \ref{2wav}, where both the {\em real} and {\em
imaginary parts} of the four measured matrix elements of the unitary
operator $U$ of the analyzed device are reported for two different
devices, and compared with the theoretical values.
As one can see, the experimental results are in very good
agreement with theory, within experimental errors. As a first
experimental demonstration we have considered only unitary devices,
however, it is clear that the method works for non unitary devices as well. 
It is also obvious that the method can be used to characterize
$n$-qubit devices---e. g. a controlled-NOT gate---in which
case we just need multiply by $n$ the whole setup, by providing an input
entangled state and two Pauli detectors for each qubit of the device,
with the full quantum characterization of the the device obtained by a
joint tomography on all output entangled pairs.
It is clear that the precision of the method will not depend on the
particular tested device---whether  it is unitary or not---and will
also be independent on the number of qubits. What makes the method
particularly reliable in the present single-photon polarization
encoding is the fact that all measurements are performed in
coincidence, making the effect of nonunit quantum efficiency of
detectors negligible, and effectively purifying the input
entangled state. In a different context---e. g. for continuous 
variables, such as homodyne tomography of twin-beams 
\cite{kumardariano}---quantum efficiency and 
entanglement purity will actually affect the final result: however,
the quantum tomographic reconstruction can handle all these kinds of 
detection noises below some thresholds\cite{varenna}, and a mixed
input state in place of $|\Psi\rangle\!\rangle$ work well 
(but less efficiently) as long as the state is {\em 
faithful}, namely it is related to a maximally entangled one by an invertible map
\cite{faith}. Unfortunately, for the twin-beam homodyne tomography
\cite{kumardariano}, faithfulness requires the knowledge of the phase
of the pump relative to the local oscillator---a feasible but difficult 
experimental task---whereas in the present experiment
the form of the entangled input state is completely under control,
being determined only by the orientation of the nonlinear crystal with
respect to the pump wave-vector and polarization.

In conclusion, we have given the first demonstration of a new
tomographic method which allows us to 
perform a complete characterization of any quantum device,  exploiting the
intrinsic parallelism of quantum entanglement, with a single entangled
state playing the role of all possible input states in quantum
parallel. The method works for any generally non unitary
multi-qubit device, and is particularly reliable in the  present
context of single-photon polarization encoding of the qubit. 

This work has been supported by the FET European Network on QIC
(Contract IST-2000-29681-ATESIT) and the INFM PRA 2001 CLON.

\end{document}